\documentclass[11pt, oneside]{article}   	
\usepackage{geometry}                		
\usepackage{graphicx}				
					          
\usepackage{amssymb}
\title{On a digital quantum description of physical space versus the continuum description}
\author{Ronald J. Adler* \\
Hansen Experimental Physics Laboratory, Gravity Probe B Mission, \\ Stanford University, Stanford, California 94309 and
\\Department of Physics and Astronomy, \\ San Francisco State University, San Francisco, California 94132}\date{February 9, 2014}				
\begin{document}
\maketitle
\begin{abstract}
The continuum of real numbers has served well as a model for physical space in mechanics and field theories. However it is a well-motivated and popular idea that at the fundamental Planck scale the combination of gravitational and quantum effects forces us to re-evaluate the concept of space (and time), and some sort of discreteness or granularity is expected. Here we study a cubic grid of points representing fundamental volumes of space and derive the spectra of momenta and wave number, which are not generally proportional to each other. The momentum is bounded and discrete for a finite grid. The uncertainty principle must be modified to reflect small scale spatial non-locality but it is consistent with the standard uncertainty principle in the continuum limit. We only consider kinematics in this work, and do not discuss time or energy or dynamical evolution. 
\end{abstract}

*electronic mail address: adler@relgyro.stanford.edu or gyroron@gmail.com \\ 
PACS numbers: fill in

\section{Introduction}

As used in physics the real numbers can be viewed as corresponding to the points on a line, and can also be viewed as an infinite string of integers.\cite{Reals} The real numbers form a continuum $\mathfrak{c}$, which is a set that cannot be put in one-to-one correspondence with the integers $\aleph_0$ and thus has more elements or a higher order infinity. The continuum has served as a model for describing physical space since the beginning of physics. It has been the basis for formulating classical mechanics and later for classical field theories such as Maxwell's electromagnetism, and still later for quantum field theories (QFT) such as quantum electrodynamics (QED) and quantum chromodynamics (QCD). Almost needless to say the model has been very successful. There is at present no experimental evidence that the continuum does not provide a valid description of space (and time) at any scale studied. For example the particle experiments at CERN's Large Hadron Collider (LHC) probe distances of order $10^{-18} - 10^{-19}m$, corresponding to a few TeV energy.\cite{LHC, Carlsmith}

However various conceptual and theoretical problems occur for $\mathfrak{c}$, and the assumption of a continuum model for physical space has been questioned by many people.\cite{Adler1, Adler2, Kleinert1, Kleinert2, Oriti1} Many of the problems are related to the infinite density of degrees of freedom inherent in the continuum. For example the self-energy of a point charge in classical electromagnetism is infinite, and the analogous divergences of QFTs are only handled by clever artifices in renormalization theory.\cite{BjD,Peskin} Of course even if no divergences occurred perturbative QFTs would require renormalization, but the divergences make the process much more delicate and difficult to reconcile with relativistic invariance. 

At the most fundamental conceptual level there is an obvious intuitive strain associated with the real number continuum: the points on a line of any length can be put in one to one correspondence with the points of any other line, so any two lines have, by definition, the same number of points. (As a trivial example we may place the points $x$ in $[0,1]$ in one-to-one correspondence with the points $y$ in $[0,2]$ simply by taking $2x=y$.) Thus a line of about Planck length $10^{-35} m$ has the same number of points as a line of about Hubble length $10^{26} m$, so the number of degrees of freedom is the same for the smallest and largest distances that typically occur in present day physics. Taken at face value this property of lines must strain our intuition of what physical space ``should" be like, and it motivates a viewpoint in which ÒpointsÓ are replaced as fundamental things by ÒelementsÓ of a different sort, that is with a discreteness or granularity. 

Special relativity does not easily accommodate such a granularity since it postulates a physical space that is intrinsically invariant under translations and rotations and other Lorentz transformations; such invariance is consistent with a continuum space but not with most models of granularity. In particular it is an attractive and popular idea that at the Planck scale spacetime should display quantum properties and be in some way granular or discrete and relativity must be modified.\cite{Adler1, Oriti2}

Theories such as string theory, loop quantum gravity, causal set theory, various spin network theories and condensed matter analog theories can produce ``emergent" spaces with various forms of discreteness. \cite{Smolin1,LQG,Oriti2, Laughlin} In this paper we will take such discreteness as given, with strong motivation, and describe space by a cubic grid of points; the grid is quite similar to the world crystal of Kleinert.\cite{Kleinert1} We prefer to think of the points in the grid as representing small fundamental regions, presumably of about the Planck volume; that is, the points form a discrete ÒparliamentÓ for regions whose detailed shape need not be specified. In this paper we use the term grid rather than lattice to avoid confusion with the Wilson lattice used in lattice gauge theory.\cite{Greensite}

In quantum mechanics (QM) the position of a particle is represented by an observable operator $X$ with dynamical properties, and we may think of the allowed positions of a particle as corresponding to the geometry of the system. Conversely in QFT space merely labels the continuous degrees of freedom of the field, with no dynamical meaning. Thus, somewhat paradoxically, it seems more profitable to study discrete versus continuum space in a QM particle context rather than a QFT context. Specifically, the operator {X} in QM is ordinarily taken to correspond to observing a particle with a position eigenvalue {x}. Here we take it to correspond to the physically allowed points or small regions in space  - the quantum geometry. 

The mathematical task of the present paper is to describe the position, wave number and momentum spectra allowed for particles and thus for space itself. With the assumption of a difference operator to replace the differential operator of standard QM the task can be accomplished analytically and exactly. An important point to emphasize is that in this paper we only analyze QM kinematics, that is the space position spectrum and the wave number and momentum spectra; we make no mention at all of time or energy or dynamics. Thus we do not analyze waves, or consider a Schrodinger type of equation for time evolution. 

In the standard continuous case the momentum and wave numbers of QM are related in one dimension by $p = \hbar k$, where $p$ and $k$ are any real numbers. In the discrete grid case this relation is altered so $p$ is not proportional to $k$. For an infinite grid both $p$ and $k$ have a limited range and are continuous; for a finite grid both $p$ and $k$ have a limited range but are discrete. The physical situation and the results are similar to those for phonons in a crystal for obvious reasons. 

Since position and momentum are modified from standard QM the uncertainty principle (UP) is also modified, but remains consistent with the usual UP in the appropriate approximation. The modified UP reflects the non-locality of the space.

Finally, from the completely different perspective of practicality, we note the fact that we routinely adopt a digital description of physical phenomena for all but the simplest calculations, since computers simply do not understand the continuum. Of course the present digitization of nearly everything is not related to any fundamental discreteness in space, but is done for practical reasons using various appropriate digitization scales and schemes.  

The paper is organized as follows. Sec.2 is a brief review of the standard QM treatment of momentum; it forms a baseline from which the discrete case is done by close analogy.  Sec.3 is an analysis of the momentum problem for an infinite grid; in order to define a momentum operator we replace the differential operator of standard QM with a difference operator. The commutator of the position and momentum operators is studied in sec.4, along with the corresponding modified UP, which we show to be consistent with the standard uncertainty relation in the appropriate approximation. In sec.5 the three dimensional problem is discussed, trivially built from the one-dimensional work. Sec.6 deals with the wave number and momentum spectra for a finite grid rather than the infinite grid of sec.3; both spectra are bounded and discrete. In sec.7 approximately the same momentum and wave number spectra are obtained using an ansatz with rather interesting properties. Sec.8 discusses a naive model of a cubic grid Universe based on the position and momentum spectrum of sec.6 and sec.7; we also summarize our results in that section, and comment on the obvious limitations of a cubic grid model of space, in particular its lack of rotational and Lorentz invariance. In appendices A and B we discuss in more detail the difference operator used to define momentum. 

\section{The standard continuum position and momentum spectra}
We first review briefly how position and momentum operators are handled in standard one dimensional QM. \cite{Shankar, Dirac1} This will be done in a fairly complete and formal way, such that the treatment of the discrete spectrum to be studied in sec.3 follows the same procedure step by step. The position operator $X$ is defined by 
\begin{equation}
\label{1}
{X|x\rangle=x|x\rangle} , \; {\langle{x' }|x\rangle=\delta({x'}-x)}.
\end{equation}
Thus the matrix elements of {X}  in the position basis are 
\begin{equation}
\label{2}
{X_{x'x} =\langle{x'}|X|x\rangle=x\delta({x'}-x)}. 
\end{equation}
It follows that for any Hilbert space vector  ${|f\rangle}$,
\begin{equation}
\label{3}
{\langle x|X|f\rangle=x\langle x|f\rangle \equiv xf(x)},
\end{equation}
where $f(x) \equiv \langle x|f \rangle$ defines the wave function. 

Similarly the derivative operator $D$  is defined to obey
\begin{equation}
\label{4}
{D|f \rangle\equiv |df/dx\rangle, \; \langle x|D|f\rangle=\langle x|df/dx\rangle=df(x)/dx},
\end{equation}
and the momentum operator is defined as $P=-i \hbar D$  so that
 \begin{equation}
\label{5}
\langle x|P|f\rangle=-i\hbar \langle df/dx\rangle=-i\hbar [df(x)/dx]. 
\end{equation}						
Note that we do not use units in which  $\hbar=1$ since we will distinguish between momentum and wave number for a discrete space; see appendix C. The matrix elements of  $P$ in the position  basis are easy to obtain using the Dirac procedure of inserting a complete set of states; from eq.(5) 
 \begin{equation}
\label{6}
-i\hbar [df(x)/dx] =\int dx' \langle x|P|x'\rangle \langle x'|f \rangle=\int dx' \langle x|P|x'\rangle f(x').
\end{equation}	
Thus the matrix elements of $P$ may be expressed as
 \begin{equation}
\label{7}
 P_{xx'} =\langle x|P|x'\rangle=-i\hbar \delta (x-x') [d/dx].
\end{equation}	
From the matrix form of the $P$ operator we then obtain a differential equation for its eigenvalues $p$ and eigenfunctions $\psi_p(x)=\langle x|p \rangle$, 
 \begin{equation}
\label{8}
-i\hbar \int dx'\delta(x-x')[d\psi_p (x')/dx']=p\psi_p (x)=-i \hbar  [d\psi_p (x)/dx].
\end{equation}	
Thus the eigenvalues $p$ are any real numbers  and the eigenfunctions in the position  basis are 
 \begin{equation}
\label{9}
\psi_p (x)=\langle x|p \rangle=\exp(ipx/\hbar)\psi_p (0) 
\end{equation}	
Eq.(9) is the complete solution for the continuous momentum eigenstates in one dimension. 

\section{Wave number and momentum spectra for an infinite grid}

We now follow a procedure analogous to that of sec.2 to obtain the wave number and momentum spectrum for a discrete space. The discrete position spectrum is defined by   
\begin{equation}
\label{10}
{X|n\rangle=n\epsilon|n\rangle} , \; \langle{n' }|n\rangle=\delta_{n'n}.
\end{equation}
where $\epsilon$ is the grid spacing and $n'$ and $n$ are integers. 
The matrix elements of $X$ are
\begin{equation}
\label{11}
X_{n'n} =\langle{n'}|X|n\rangle=n \epsilon \delta_{n'n}, 
\end{equation}
and analogous to eq.(3) we have
\begin{equation}
\label{12}
\langle{n}|X|f\rangle=n \epsilon \langle n|f \rangle=n\epsilon f(n).
\end{equation}
 								
Next we replace the derivative operator $D$ in eq.(4) with a difference operator $D\rightarrow \Delta / \epsilon$, defined by the analog of eq.(4),
\begin{equation}
\label{13}
(\Delta / \epsilon) |f \rangle=(1/\epsilon)|\Delta f\rangle, \; \langle n|(\Delta / \epsilon )|f\rangle=(1/\epsilon) \Delta f(n), 
\end{equation}
where the difference operator $\Delta$ is defined as\begin{equation}
\label{14}
 \Delta f(n)=(1/2)[f(n+1)-f(n-1)].
\end{equation}
The motivation for this choice of a difference operator is rather compelling as we will discuss below. Choosing the momentum operator in the obvious way as $P\rightarrow -i\hbar(\Delta / \epsilon)$ we obtain the analog of eq.(5), 
\begin{equation}
\label{15}
\langle n|P|f \rangle =(-i \hbar)[f(n+1)-f(n-1)]/2 \epsilon.
\end{equation}
Next we insert a complete set of states in eq.(15) to obtain a matrix equation 
\begin{equation}
\label{16}
\sum\limits_{n'}
\langle n|P|n' \rangle f(n')=(-i \hbar / 2\epsilon)[f(n+1)-f(n-1)].
\end{equation}
From the last expression we may extract the matrix elements of $P$ in the discrete position basis, 
\begin{equation}
\label{17}
P_{nn'}=\langle n|P|n' \rangle =(-i \hbar / 2\epsilon)[\delta_{n+1,n'}-\delta_{n-1,n'}],
\end{equation}
which is the analog of eq.(7). Note that the $P$ matrix is Hermitian, which justifies our choice of the difference operator in eq.(14). See also appendices A and B. 

The continuum eigenvalue eq.(8) is replaced in the discrete case by the difference equation
\begin{equation}
\label{18}
\psi_p (n+1)-\psi_p(n-1)=(2ip/\epsilon)\psi_p (n).
\end{equation}
The solution to this difference equation is analogous to that of the differential eq.(8). We use an exponential ansatz with a wave number $k$  to be determined, 
 \begin{equation}
\label{19}
\psi_p (n)=\exp(ikn\epsilon),
\end{equation}	
and find by substitution that the wave number and momentum are related by
 \begin{equation}
\label{20}
p \epsilon / \hbar = \sin (\epsilon k).
\end{equation}	
Fig.1 shows the relation between the wave number $k$ and momentum $p$ in eq.(20); it is what one expects by analogy with phonons in a crystal. In the limit of small $\epsilon$  eq.(20) gives the continuum result $p=\hbar k$. In general eq.(20) tells us that $p$ can be any real number between $-\hbar / \epsilon$ and $\hbar / \epsilon$, with the wave number $k$ given by $(1/\epsilon) sin^{-1}(p\epsilon/ \hbar)$. That is, the momentum and wave number are not proportional, contrary to the continuum result. See Appendix C.

\section{The $X,P$ commutator and the uncertainty principle}
The discrete position matrix $X$ is diagonal by demand, with integer elements that run from negative infinity to positive infinity, while the momentum matrix $P$ in eq.(17) has $+1$ elements on the super-diagonal and $-1$ elements on the sub-diagonal; they are explicitly 
\begin{equation}
\label{21}
X=\epsilon \left( \begin{array}{ccccc} ... & ... & ... & ... & ...\\ ... & 1 & 0 & 0 & ... \\ ... & 0 & 2 &  0 & ...\\ ... & 0 & 0 & 3 & ... \\  ... & ... & ... & ... & ...\end{array} \right),  \;
P=(-i \hbar / 2\epsilon )\left( \begin{array}{ccccc} ... & ... & ... & ... & ...\\ ... & 0 & 1 &  0 & ... \\ ... & -1 & 0 & 1 & ...\\ ... & 0 & -1& 0 & ... \\ ... & ... & ... & ... & ...\end{array} \right).   
\end{equation}
The matrix $P$ is Hermitian due to our choice of the difference operator $\Delta$ in eq.(14). That choice is quite compelling; for example, if we had chosen the difference operator to be the simpler difference $f(n+1)-f(n)$ then $P$ would have all ones on the super-diagonal and negative ones on the diagonal, and would not be Hermitian. In appendices A and B we discuss the choice of the difference operator further.
 		
As a result of our choice of the difference operator the commutator of $X$ and $P$ has $+1$ on both the super-diagonal and sub-diagonal, and is explicitly
\begin{equation}
\label{22}
[X,P]=(i \hbar /2) \left( \begin{array}{ccccc} ... & ... & ... & ... & ...\\ ... & 0 & 1 & 0 & ...\\ ... & 1 & 0 & 1 & ...\\ ... & 0 & 1 & 0 & ...
\\ ... & ... & ... & ... & ...\end{array} \right)  = i\hbar I_\Delta .
\end{equation}
The off diagonal elements correspond to a basic non-locality in space: neighboring points do not correspond to independent degrees of freedom. Note also that the trace of both sides of eq.(22) is zero, so the zeros on the diagonal of the matrix $I_\Delta$ are to be expected. The same commutator is obtained in appendix A, eq.(50), by somewhat different means. 

From the commutator in eq.(22) we may obtain an uncertainty principle (UP) by the standard QM procedure.\cite{Shankar} The result is  
\begin{equation}
\label{23}
\Delta X \Delta P \geq (1/2)|\langle [X,P]/i \rangle| =( \hbar /2) \langle I_\Delta \rangle.
\end{equation}	
This is consistent with the standard UP as can be seen as follows. We denote the elements of the state used in forming the expectation value in eq.(23) as $v_j$ so eq.(23) becomes
 \begin{equation}
\label{24}
\Delta X \Delta P \geq (\hbar/2)\sum_{j} v_{j }^{*}(v_{j+1} + v_{j-1})/2.
\end{equation}	
On a scale large compared to $\epsilon$ we expect the elements $v_j$ to vary little over the small distance $\epsilon$ and thus be slowly varying functions of $j$; then $(v_{j+1} + v_{j-1})/2 \simeq v_j$. Taking the $v_j$ to be normalized we thus see that the UP becomes  \begin{equation}
\label{25}
\Delta X \Delta P \geq (\hbar /2)\sum_{j} v_{j}^{*} v_j=\hbar/2 ,
\end{equation}	
which is the standard UP. 

\section{Three dimensions}
The transition to three dimensions is quite straight-forward and only requires brief comments. For example the continuum normalization eq.(1) becomes  
\begin{equation}
\label{26}
\langle \vec x'| \vec x \rangle = \delta (\vec x'- \vec x) \delta (\vec y'-\vec y)\delta (\vec z'-\vec z)
\end{equation}	
and the eigenvalue eq.(8) and its solution become
\begin{equation}
\label{27}
\nabla \psi_p (x)=(i \vec p / \hbar ) \psi_p ( \vec x),  \space \langle \vec x| \vec p \rangle 
=\psi_p (\vec x) = exp(i \vec p \cdot \vec x /\hbar) \psi_p(0).
\end{equation}	

For the discrete case the integer $n$  is replaced with a three-vector of integers, so
\begin{equation}
\label{28}
\vec n=(n_1 , n_2, n_3), \; \vec x = \vec n \epsilon=\vec n=(n_1 , n_2, n_3).
\end{equation}	
The spatial spectrum is an infinite cube of integers. The difference operator eq.(14) becomes a vector difference operator
\begin{equation}
\label{29}
\Delta_j f( \vec nj=(1/2)[f( \vec n + \vec u_j)-f( \vec n - \vec u_j)] , 
\end{equation}	
where $ \vec u_1 = (1,0,0,0)$ and similarly for the 2 and 3 components.  Most important the eigenfunction in eq.(19) and the relation eq.(20) between the wave number and momentum become
\begin{equation}
\label{30}
\psi_p ( \vec n)=\exp(i \vec k \cdot \vec n \epsilon)\psi_p (0), \; \sin(\epsilon k_j)=p_j \epsilon / \hbar.
\end{equation}	

\section {Discrete wave number and momentum spectrum for a finite grid}
In sec.3 we looked at a discrete space grid running over an infinite range and found that the momentum spectrum is continuous but has a finite range, from $- \hbar/\epsilon$ to $ \hbar/\epsilon$; this is to be expected since the wave length cannot be smaller than the grid spacing  $\epsilon$. In this section we will  study a discrete space spectrum running over a finite range and obtain the wave number and momentum spectra, which turn out to be discrete. Thus we consider matrices  $X$ and $P$ which are truncated versions of eq.(21), and for $N=2$ are the following 
\begin{equation}
\label{31}
X=\epsilon \left( \begin{array}{ccccc} -2 & 0 & 0 & 0 & 0\\ 0 & -1 & 0 & 0 & 0 \\ 0 & 0 & 0 &  0 & 0\\ 0 & 0 & 0 & 1 & 0 \\  0 & 0 & 0 & 0 & 2\end{array} \right),  \;
P=(-i \hbar / 2\epsilon )\left( \begin{array}{ccccc} 0 & 1 & 0 & 0 & 0\\ -1 & 0 & 1 &  0 & 0 \\ 0 & -1 & 0 & 1 & 0\\ 0 & 0 & -1& 0 & 1 \\ 0 & 0 & 0 & -1 & 0 \end{array} \right).   
\end{equation}
The momentum spectrum thus corresponds to the zeroes of an $M=2N+1$ order characteristic polynomial. For the present case this is quite tractable and can be done in closed form. 
	
With the definition $\lambda = 2p \epsilon / \hbar$ the characteristic polynomial can be written, displaying the case $M=5$, as 
\begin{equation}
\label{32}
det \left( \begin{array}{ccccc} \lambda & i & 0 & 0 & 0\\ -i & \lambda & i &  0 & 0 \\ 0 & -i &\lambda & i & 0\\ 0 & 0 & -i & \lambda  & i \\ 0 & 0 & 0 & -i & 0 \end{array} \right) = D_5(\lambda)=0,   
\end{equation}
and so forth for any $M$. It is easy to see that the polynomials $D_M(\lambda)$, defined as in eq.(31), obey the recursion relation
\begin{equation}
\label{33}
D_M(\lambda)=D_{M-1}(\lambda)-D_{M-2}(\lambda).   
\end{equation}
The first three $D_M(\lambda)$ are
\begin{equation}
\label{34}
D_0(\lambda)=1, D_1(\lambda)=\lambda, D_2(\lambda)=\lambda ^2 -1. 
\end{equation}
The recursion relation eq.(33) with the first two polynomials given in eq.(34) is that of the Chebyshev polynomials of the second kind, $U_M(x)$, if we identify the argument as $x=\lambda /2$.\cite{Abromowitz,Cheb} Thus to get the eigenvalues of the $P$  matrix we need the zeros of the Chebyshev $U_M(x)$. 

The zeros are most easily obtained with the use of the identity\cite{Cheb}
\begin{equation}
\label{35}
U_M(\cos\theta)=\frac{\sin(M+1)\theta}{\sin\theta}. 
\end{equation}	
Thus the zeros are at
\begin{equation}
\label{36}
\theta=\frac{l\pi}{M+1}, \lambda = 2 \cos \left( \frac{l\pi}{M+1}\right),
\end{equation}	
where $l$ is an integer. The momentum eigenvalues are therefore give by
\begin{equation}
\label{37}
\frac{p_l \epsilon}{\hbar} = \cos\left( \frac{l \pi}{2(N+1)} \right).
\end{equation}	
To compare the spectrum in eq.(37) with the continuous momentum spectrum in eq.(20) we relabel the eigenvalues with $j=(N+1)-l$  and get
\begin{equation}
\label{38}
\frac{p_j \epsilon}{\hbar} =\sin \left( \frac{\pi j}{2(N+1)}  \right).
\end{equation}	
From the last expression it is clear that for positive $j$  the maximum wave number is $|k_{max}\epsilon |=\pi /2$  and the maximum momentum is $|p_{max}\epsilon |=\hbar /\epsilon$ for $j=N+1$, and similarly for negative momenta. 
The agreement of this with the continuous expression in eq.(20) is obvious and pleasing. See fig.1. 

The momentum spectrum in eq.(38) is the complete solution to the problem in one dimension, and the solution for three dimensions is obvious from the discussion of sec.5.

\section{An alternative ansatz for the discrete momentum spectrum}

In this section we will present a somewhat curious ansatz that makes the problem of the discrete momentum spectrum very easy for large values of $N$. It is an alternative to the approach of sec.6. 

We make the basic assumption that the effect of truncating the space grid does not greatly affect the momentum spectrum and are led to adopt the standard artifice of imposing periodic boundary conditions on the momentum function $\psi(n)$. Specifically, we retain the exponential solution in eq.(19) but demand that 
\begin{equation}
\label{39}
\psi(N)=\pm \psi(-N).
\end{equation}
The $ \pm$ is important as we will see. This boundary condition naturally results in a discrete wave number spectrum. 

Note that $\psi(n)= \psi ^* (-n)$ from eq.(19). Thus for the choice of the plus sign in eq.(39) we see that $\psi(N)=\psi ^* (N)$  is real, and therefore
\begin{equation}
\label{40}
\exp(ikN \epsilon)=\cos(kN\epsilon)+i\sin(kN\epsilon)=real, \; kN\epsilon=0, \pm\pi, \pm 2\pi, ... .
\end{equation}
For the choice of the minus sign in eq.(39) we similarly see that 
\begin{equation}
\label{41}
\exp(ikN \epsilon)=\cos(kN\epsilon)+i\sin(kN\epsilon)=imaginary, \; kN\epsilon=\pm\pi /2, \pm 3\pi /2, ... .
\end{equation} 
Combining the two cases, eq.(40) and eq.(41),  we see that 
\begin{equation}
\label{42}
kN\epsilon=0, \pm\pi /2, \pm \pi , ... .
\end{equation} 
Thus the momentum spectrum changes from continuous in eq.(20) to the discrete
\begin{equation}
\label{43}
p_j \epsilon / \hbar = \sin(\pi j /2N),
\end{equation} 
where $j$ runs from $-N$ to $N$; eq.(43) is the same as the exact result eq.(38) but with $N$ replacing $N+1$. 
See fig.1. 
 
The above procedure is equivalent to thinking of the position $x$  spectrum as residing on a circle or one dimensional torus; in three dimensions it corresponds to thinking of the $ \vec x$  spectrum as being on a flat 3 torus. The $ \pm$ boundary condition is necessary and rather curious and reminiscent of the multivalued behavior exhibited by spinors.
             
In summary, the momentum eigenvalues for large $N$ are given by eq.(38) from the exact analysis in sec.6, or eq.(43) from the ansatz above. We find it interesting that the ansatz provides the correct expression only if both the plus and minus boundary conditions are used, but we have at present no physically intuitive explanation for this. 

\section{A na\"{i}ve cubic grid Universe}
In this section we take $N$ to be very large and thus $M=2N$. The na\"{i}ve three-space we have studied is a cubic grid with $M$ points on a side and spacing $\epsilon$; its size is $L=M\epsilon=2N \epsilon$.  Our preferred interpretation is that each point in the cube represents a region of volume $\epsilon ^3$, rather than a single point in space. That is the space has a granular structure without a specified shape, but with an intrinsic minimum meaningful distance $\epsilon$. 

An obvious natural choice for  the spacing is the Planck distance $L_P =1.6 \times 10^{-35} m$, which is widely considered to be the scale at which the concept of distance must be reconsidered to include gravitational and quantum effects.\cite{Adler1,Oriti1} We suggest that the distance $10^{-35} m$ be called a Planck, by analogy with the fermi $10^{-15}m$, and will do so henceforth. The size of the cube we take to be about twice the Hubble distance $L_H=1.2 \times 10^{26} m$, which is also approximately the deSitter radius.\cite{ Adler1,Adler2,Liddle} Thus the number $M$ is 
\begin{equation}
\label{44}
M=L/ \epsilon = 2L_H/\epsilon =1.5 \times 10^{61}. 
\end{equation} 
It has been suggested that the inverse of this number be identified as a ``gravitational fine structure constant" $\alpha_g\approx 10^{-61}$.\cite{Adler3}

Similarly, according to eq.(42) and eq.(43) the wave number space is a cubic grid with spacing $\Delta k=\pi /2N\epsilon$ and $M=2N$ points on a side; the index $j$ runs from $-N$ to $N$. Thus the wave number spacing is about $1.3 \times 10^{-26}m^{-1}$. The maximum wave number is $k_{max}=\pi/ 2\epsilon$ or about $9.8 \times 10^{34}m^{-1}$. The wavelengths corresponding to these wave number are $\lambda _{max}=4L_H = 4.8 \times 10^{26}m$ and $\lambda_{min} = 4\epsilon=6.4 \times 10^{-35}m$, which is what one obviously should expect. Fig.2 shows the cubic grids for the space spectrum and the wave number spectrum. 

The momentum space, according to eq.(42), is a cube of $M=2N$ point on a side, but the spacing is not uniform. For small $j$ and wave number $k$ and momentum $p$ we have approximately
\begin{equation}
\label{45}
p_j = (\hbar/\epsilon)\sin(\pi j/2N) \simeq \hbar k_j,
\end{equation} 
as expected. For the largest allowed $j=N$ the momentum is maximum 
\begin{equation}
\label{46}
p_{max} = \hbar/\epsilon, 
\end{equation} 
which is the Planck energy $1.2 \times 10^{19} GeV$ over c. The spacing of the momentum eigenvalues goes to zero at the maximum. 

It is readily apparent that the cubic model Universe is na\"{i}ve because of its obvious limiting features: (1) it is not relativistically covariant since time is not included; (2) it is not rotationally invariant since it is cubic; (3) it does not include gravitational curvature effects since it is flat. However the model is not quite as limited as it may appear, as pointed out by Kleinert.\cite{Kleinert1} This is because: (1) the model may be extended in an obvious way to a Euclidean 4-hypercube with imaginary time, with use of analytic continuation to restore the appropriate signature; (2) various theoretical considerations such as the generalized uncertainty principle discussed in reference 21 suggest large intrinsic quantum fluctuations at the Planck scale, which would wash out the cubic crystal structure and make the cube effectively amorphous on any scale, hence rotationally and Lorentz invariant; (3) large scale gravitational curvature effects may be viewed as crystal defects called disclinations, consistent with general relativity.\cite{Kleinert1,Kleinert2} 

\section{Acknowledgements}

We thank the Stanford Gravity Probe B theory group for discussions and particularly Paul Worden, Robert Wagoner and James Bjorken for stimulating comments about fundamental space and time structure. Hagen Kleinert provided helpful insightful comments on the relation of this work to his world crystal model and the description of gravitational curvature by crystal defects. 
\appendix
\section{Some properties of the difference operator}

The difference operator $\Delta$ in eq.(14) is Hermitian and has some interesting properties. For example he analog of the product rule for derivatives is 
\begin{equation}
\label{47}
\Delta \left ( f(n)g(n) \right ) = \bar{f} \Delta g+\bar {g}\Delta f,
\end{equation} 
where 
\begin{equation}
\label{48}
\bar{f} \equiv \left ( f(n+1)+f(n-1) \right )/2, 
\end{equation} 
and similarly for $g$. For the special case of $g(n)=n$ this gives
\begin{equation}
\label{49}
\Delta \left ( f(n)n \right ) =2 \bar{f} +n\Delta f.
\end{equation} 
From the last relation we may calculate the commutator $[X,P]$ in a different way than in the text.  In the  $X$ representation the commutator operating on a function $f$ gives 
\begin{equation}
\label{50}
[X,P] f(n) =(-i \hbar/2\epsilon)\left ( n\epsilon \Delta f - \Delta (n\epsilon f) \right )=(-i \hbar/2)\left ( n \Delta f - 2 \bar{f} - n\Delta f \right ) = i \hbar \bar{f},
\end{equation} 
which is equivalent to eq.(22). 

\section{Other difference operators}
The difference operator introduced in eq.(14) depends on the function $f$ at $n+1$ and $n-1$. It is arguably the simplest reasonable and consistent Hermitian operator. However it is not unique. It can be viewed as resulting from fitting a second order polynomial to a function specified at three points, which allows calculation of the derivative at the center point. 

One might instead fit a forth order polynomial to a function specified at 5 points and calculate its derivative at the center point. This procedure gives the difference operator  			
\begin{equation}
\label{51}
\Delta f(n) =(4/3)\left ( f(n+1)-f(n-1) \right )/2 - (1/3)\left ( f(n+2)-f(n-2) \right )/4, 
\end{equation}
which we may write in a suggestive way as
\begin{equation}
\label{52}
\Delta f(n) \equiv (4/3)\Delta_1 f- (1/4) \Delta_2 f.
\end{equation}
The corresponding  matrix contains two nonzero super-diagonals sub-diagonals, containing $4/3$ and $-1/4$. This process could be used for any number of points and would produce an antisymmetric matrix similar to that in eq.(31) with zeros on the diagonal and nonzero elements near the diagonal. We will not pursue such considerations further but will adopt the simplest option as discussed in sec.3.  

\section{Alternative physical interpretation of the momentum wave number relation}
Consider the relation eq.(20) between the momentum and wave number, or its discrete version in eq.(43). We rewrite it as 
 \begin{equation}
\label{52}
p=\hbar k \left [ \sin (\epsilon k)/\epsilon k \right] \equiv k \hbar(k,\epsilon) = k \hbar (0) \left [1-(\epsilon k)^2 /3! ... \right ].
\end{equation}
From eq.(53) we see that we may think of $\hbar$  as a function of $k\epsilon$ rather than a true constant. The relation between the momentum $p$ and the wave number $k$ can, in principle, be tested: the momentum may be measured kinematically, while the wave number may be measured via interference or diffraction effects. Almost needless to say any such measurement would be expected to be quite difficult at the relevant scale. On the other hand the distinction may be relevant for virtual processes and thereby indirectly testable.\cite{Adler5} We note also that it is not straight-forward to make the relation eq.(53) consistent with relativistic covariance.

\begin{figure}[htbp] 
  \centering
   \includegraphics[width=4in]{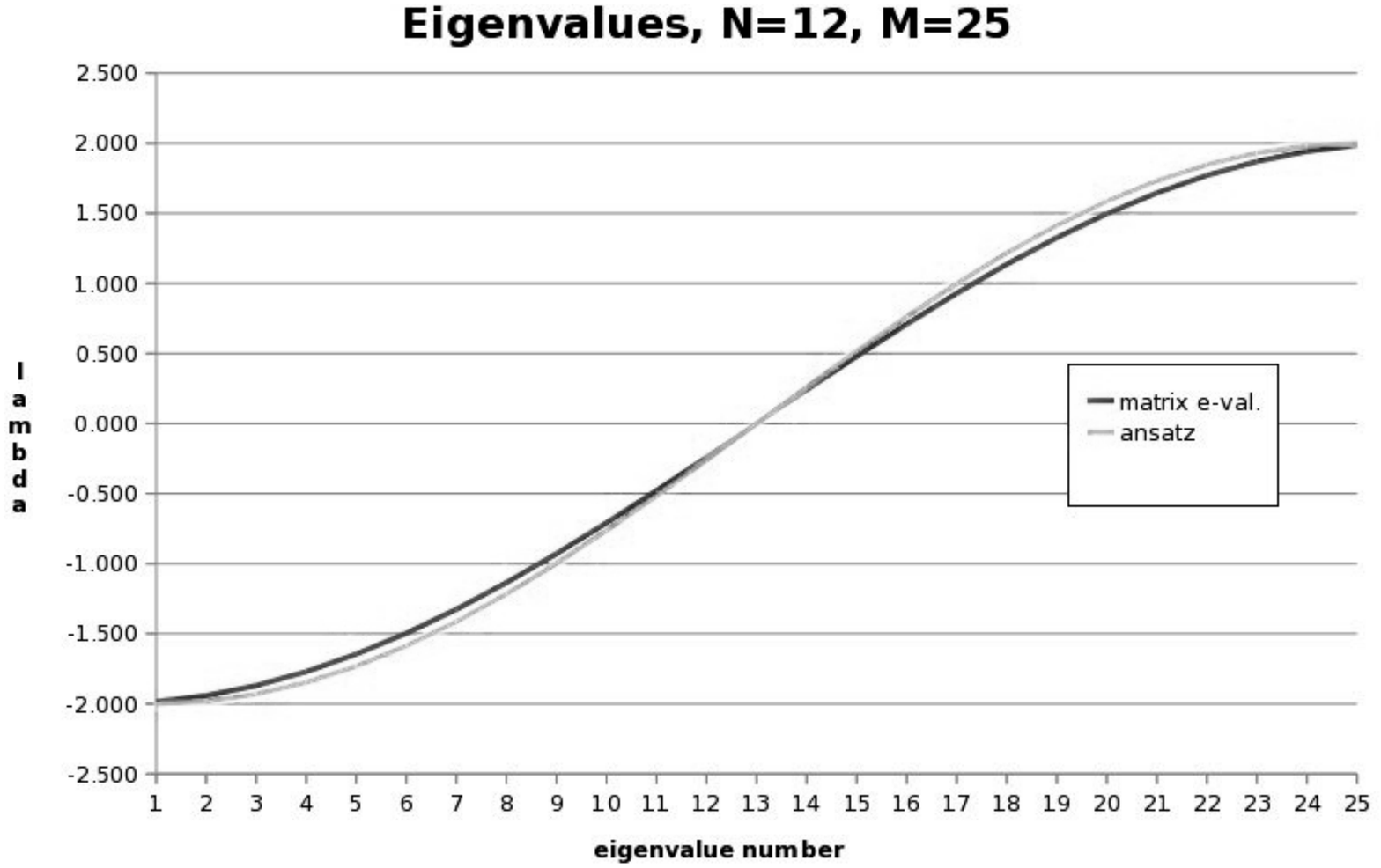} 
   \caption{Momentum ($p\epsilon/\hbar$) versus wave number index $j$ for $N=12$. The dark curve is the exact result and the light curve is the approximate result for the ansatz of sec.7.}
   \label{fig1}
\end{figure}
\begin{figure}[htpb] 
     \centering
     \includegraphics[width=5in]{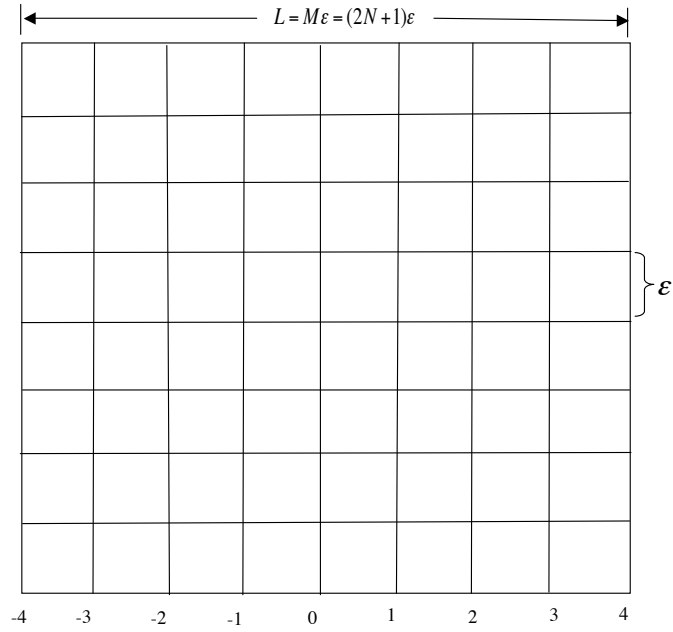} 
   \caption{End view of the na\"{i}ve Universe for $N=4$. We have in mind $N$ of order $10^{61}$.}
   \label{fig2a}
\end{figure}
\begin{figure}[htbp] 
     \centering
     \includegraphics[width=5in]{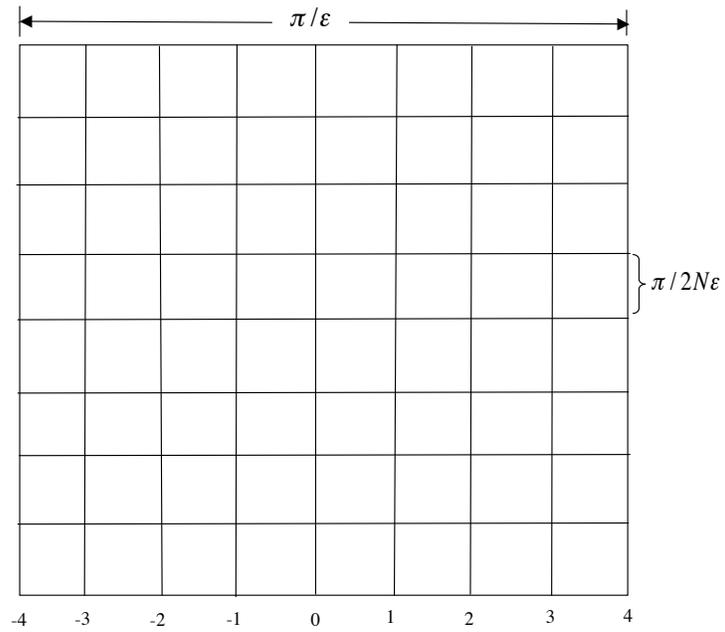} 
   \caption{End view of the na\"{i}ve Universe wave vector space.}
   \label{fig2b}
\end{figure}

 

\end{document}